\documentclass[reprint,superscriptaddress, aps]{revtex4-1}

\usepackage{blindtext}
    \usepackage{amsmath}%,amssymb}
    \usepackage{makeidx}
    \usepackage{amsfonts}
    \usepackage{float}
    \usepackage[usenames,dvipsnames]{pstricks}
    \usepackage{subfigure}
    \usepackage{epsfig}
    \usepackage{pst-grad} % For gradients
    \usepackage{pst-plot} % For axes
    \usepackage[colorlinks,hyperindex]{hyperref}
    \hypersetup
    {
        colorlinks,%
        citecolor=black,%
        linkcolor=black,%
        urlcolor=black,%
    }

\usepackage{color}
\usepackage{dsfont}
\usepackage{graphicx}% Include figure files
\usepackage{subfigure}

\usepackage{dcolumn}% Align table columns on decimal point
\usepackage{amsmath}    % need for subequations
\usepackage{verbatim}   % useful for program listings
\usepackage{color}      % use if color is used in text
\usepackage{hyperref}   % use for hypertext links, including those to external documents and URLs
\usepackage{amsfonts}
\usepackage{braket}
\usepackage{bm}% bold math
\usepackage{epstopdf}
\usepackage{titlesec}
\usepackage{capt-of}
\usepackage{float}
\begin{document}

%\preprint{Toshiba Research Europe Limited/Confidential}

\title{
Controllable photonic time-bin qubits from a quantum dot
}

\author{J. P. Lee}\,
\email{james.patrick.lee.47@gmail.com}
\affiliation{Toshiba Research Europe Limited, Cambridge Research Laboratory,\\
208 Science Park, Milton Road, Cambridge, CB4 0GZ, U.K.}
\affiliation{Engineering Department, University of Cambridge,\\
9 J. J. Thomson Avenue, Cambridge, CB3 0FA, U.K.}

\author{L. M. Wells}\,
\affiliation{Toshiba Research Europe Limited, Cambridge Research Laboratory,\\
208 Science Park, Milton Road, Cambridge, CB4 0GZ, U.K.}
\affiliation{Cavendish Laboratory, Cambridge University,\\
J. J. Thomson Avenue, Cambridge, CB3 0HE, U.K.}

\author{B. Villa}\,
\affiliation{Toshiba Research Europe Limited, Cambridge Research Laboratory,\\
208 Science Park, Milton Road, Cambridge, CB4 0GZ, U.K.}
\affiliation{Cavendish Laboratory, Cambridge University,\\
J. J. Thomson Avenue, Cambridge, CB3 0HE, U.K.}

\author{S. Kalliakos}
\affiliation{Toshiba Research Europe Limited, Cambridge Research Laboratory,\\
208 Science Park, Milton Road, Cambridge, CB4 0GZ, U.K.}

\author{R. M. Stevenson}
\affiliation{Toshiba Research Europe Limited, Cambridge Research Laboratory,\\
208 Science Park, Milton Road, Cambridge, CB4 0GZ, U.K.}

\author{D. J. P. Ellis}
\affiliation{Toshiba Research Europe Limited, Cambridge Research Laboratory,\\
208 Science Park, Milton Road, Cambridge, CB4 0GZ, U.K.}

\author{I. Farrer}
\thanks{Current affiliation: Department of Electronic \& Electrical Engineering, University of Sheffield, Mappin Street, Sheffield, S1 3JD, U.K. }
\affiliation{Cavendish Laboratory, Cambridge University,\\
J. J. Thomson Avenue, Cambridge, CB3 0HE, U.K.}

\author{D. A. Ritchie}
\affiliation{Cavendish Laboratory, Cambridge University,\\
J. J. Thomson Avenue, Cambridge, CB3 0HE, U.K.}

\author{A. J. Bennett}
\thanks{Current affiliation: Institute for Compound Semiconductors, Cardiff University, Queen's Buildings, 5 The Parade, Roath, Cardiff, CF24 3AA, U.K.}
\affiliation{Toshiba Research Europe Limited, Cambridge Research Laboratory,\\
208 Science Park, Milton Road, Cambridge, CB4 0GZ, U.K.}

\author{A. J. Shields}
\affiliation{Toshiba Research Europe Limited, Cambridge Research Laboratory,\\
208 Science Park, Milton Road, Cambridge, CB4 0GZ, U.K.}

\date{\today}%

\begin{abstract}
Photonic time bin qubits are well suited to transmission via optical fibres and waveguide circuits. The states take the form $\frac{1}{\sqrt{2}}(\alpha \ket{0} + e^{i\phi}\beta \ket{1})$, with $\ket{0}$ and $\ket{1}$ referring to the early and late time bin respectively. By controlling the phase of a laser driving a spin-flip Raman transition in a single-hole-charged InAs quantum dot we demonstrate complete control over the phase, $\phi$. We show that this photon generation process can be performed deterministically, with only a moderate loss in coherence. Finally, we encode different qubits in different energies of the Raman scattered light, demonstrating wavelength division multiplexing at the single photon level.
\end{abstract}

\maketitle

\section{Introduction}
%Single photon sources have applications in quantum communication \cite{duan2001long}, quantum relays \cite{collins2005quantum} and quantum computing \cite{knill2001scheme}.

Quantum dots (QDs) have unparalleled brightness as single photon sources and can be embedded in a variety of semiconductor devices and microcavity structures \cite{maier2014bright, pelton2002efficient}. Until recently, the qualities of the photons generated from quantum dots have lagged behind other sources such as trapped atoms and ions, which enable the creation of photons with high indistinguishabilities and controllable temporal profiles via stimulated Raman transitions \cite{kuhn2002deterministic, beugnon2006quantum, vasilev2010single}. However, in the past few years researchers have demonstrated almost perfectly indistinguishable photons from a resonantly excited QD \cite{he2013demand}, control over the spectrum of resonantly scattered light \cite{matthiesen2012subnatural, matthiesen2013phase} and the filtering of the phonon sideband and improvement of photon coherence through the use of micropillar cavities \cite{bennett2016cavity,senellart2017high}.

Inspired by atomic physics, the use of Raman scattering in quantum dots has been used to demonstrate photon energy tuning \cite{sweeney2014cavity}, generation of photons tailored for interfacing with a quantum memory \cite{beguin2017demand}, picosecond shaping of single photons \cite{pursley2018picosecond} and generation of photons coherently superposed across multiple time bins \cite{lee2018multi}

In this work we use cavity-enhanced Raman scattering to generate a single photon time-bin-encoded qubit superposed across two time bins. We show that modulating the phase difference between the driving laser pulses results in the modulation of the phase difference between the time bins of the generated single photon state, enabling complete control of a time bin qubit without the use of an interferometer. Next, we show that the coherence between the two time bins remains when the Raman transition is driven deterministically at higher laser powers. Finally, we use two driving lasers detuned to either side of the Raman transition, we encode a different time bin qubit with each laser and show that spectral filtering enables us to recover the encoded state for each frequency.

\section{Setup}

\begin{figure*}[t]
\includegraphics[width=17 cm]{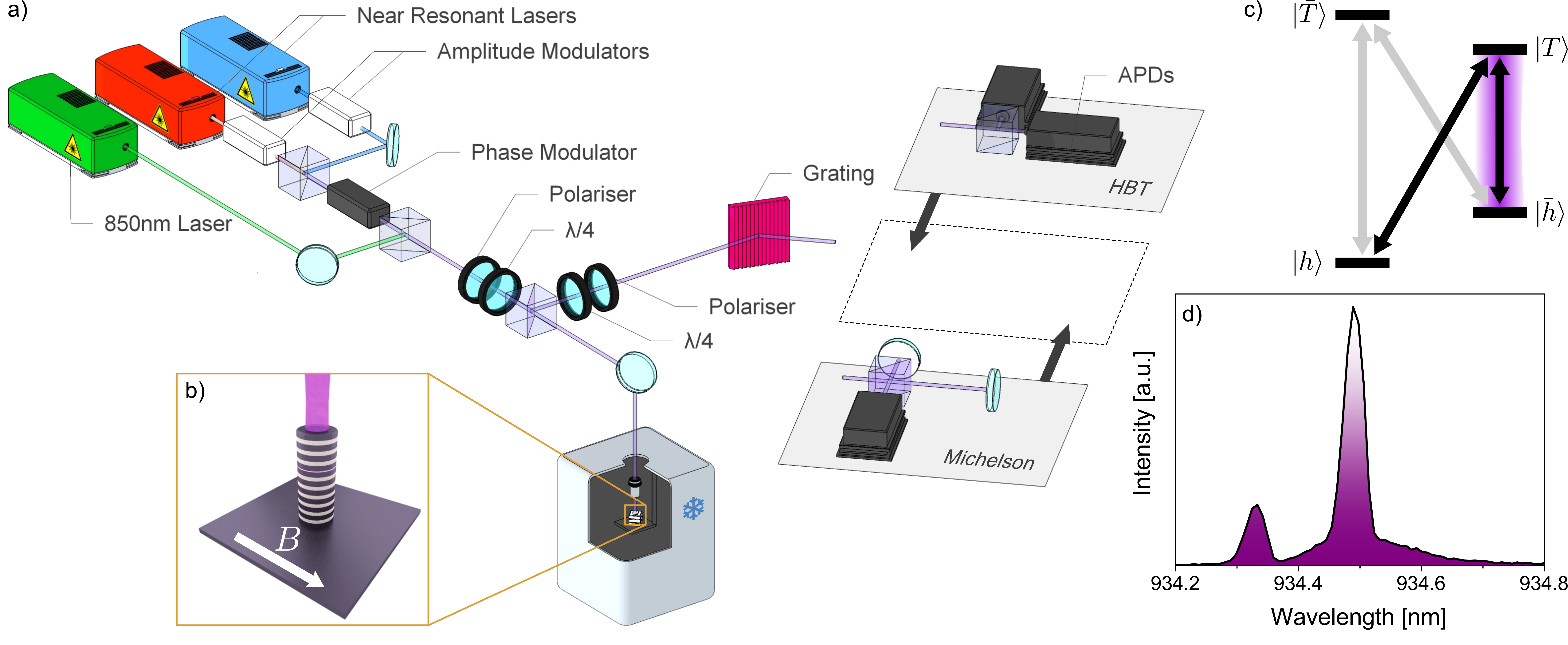}
\caption{ 	 
{\bf a)} An illustration of out experimental setup. {\bf b)} An illustration of a micropillar cavity in a Voigt geometry magnetic field. {\bf c)} The energy level diagram. When the resonant lasers are red or blue detuned from the diagonal transition, the resulting Raman scattered light is similarly detuned. {\bf d)} The spectrum of the QD under non-resonant excitation. Polarisation filtering ensures that only the vertical transitions are visible and the long wavelength transition is clearly enhanced by the cavity mode.}  
\label{Fig1}
\end{figure*}

Our experimental setup is illustrated in Figure \ref{Fig1}a. We use a single-hole-charged InAs QD held in a Voigt geometry magnetic field which results in a double lambda system, as illustrated in Figure \ref{Fig1}c. We use a narrow linewidth laser (or two narrow linewidth lasers for the wavelength division multiplexing experiment) to drive the diagonal $\ket{h}\rightarrow\ket{\bar{T}}$ transition. We use amplitude and phase modulators to control the excitation laser light. A micropillar cavity (Figure \ref{Fig1}b) is used to increase the collection efficiency and to selectively Purcell enhance emission from the longest wavelength transition. As well as the Raman scattered light having a longer wavelength than the input light, it is also orthogonally polarised \cite{warburton2013single} which allows us to use polarisation and spectral filtering to separate the driving laser light from the emitted light. The sample we use is nominally undoped, so we use a pulsed non-resonant laser to generate charge carriers in the sample in order to introduce a hole spin \cite{bennett2016cavity,bennett2016semiconductor}.
The spectrum under non-resonant excitation is shown in Figure \ref{Fig1}d.

\section{Arbitrary time bin qubits}
\subsection{Phase modulation}

\begin{figure}[t!]
\begin{minipage}[h]{0.4\textwidth}
\includegraphics[width=7 cm]{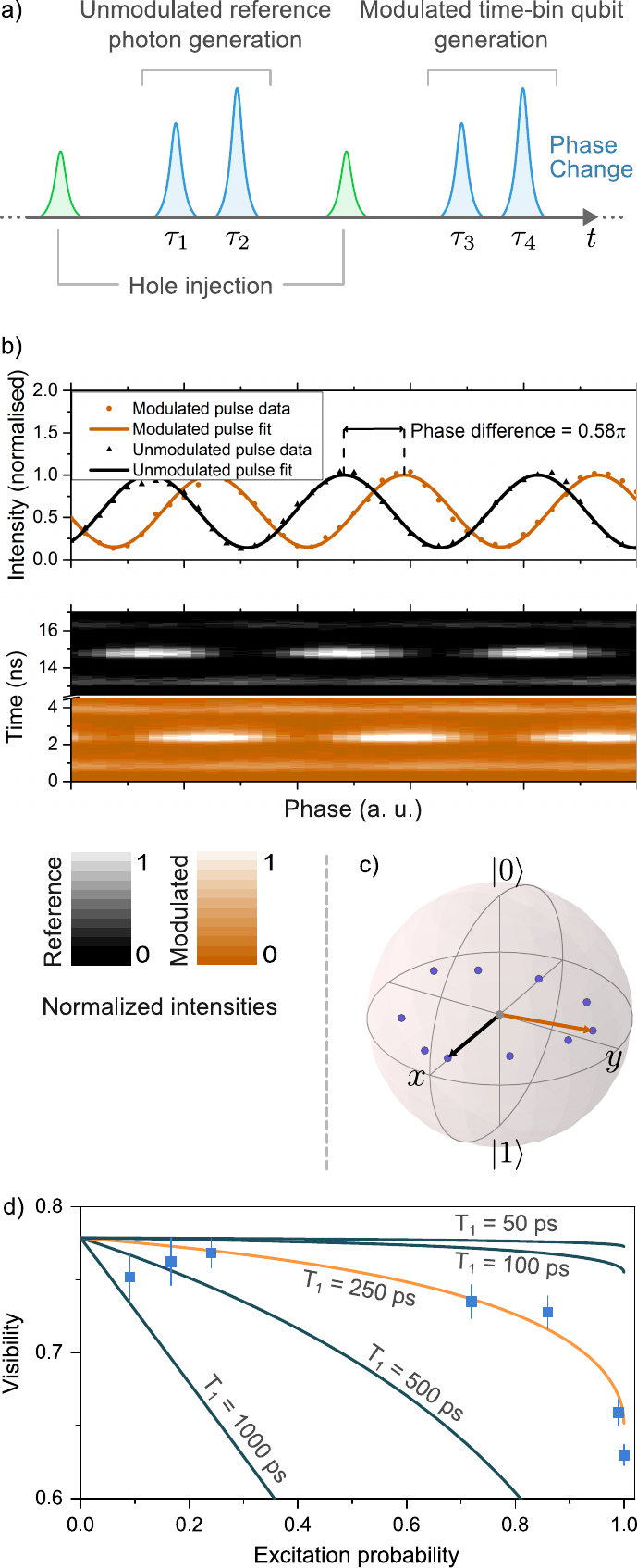}
\end{minipage}
% \begin{minipage}[h]{0.4\textwidth}
% \includegraphics[width=7 cm]{phase_mod.png}
% \end{minipage}
% \begin{minipage}[h]{0.4\textwidth}
% \includegraphics[width=5 cm]{Bloch_Sphere.png}
% \end{minipage}
% \begin{minipage}[h]{0.4\textwidth}
% \includegraphics[width=7 cm]{Coherence.png}
% \end{minipage}
\caption{{\bf a)} An illustration of the pulse sequence used to demonstrate phase control over a time bin qubit. {\bf b)} The extracted interference fringes for reference (black) and phase modulated (orange) photons. The time resolved plots of the interference between  neighbouring time bins for the two sequences that allow us to extract the interference fringes and the probabilities of measuring the photon in each time bin.   {\bf c)} The recorded time bin qubits mapped onto the Bloch sphere (blue points). The black and orange vectors represent the states generated by reference and example modulated pulse sequences respectively. {\bf d)} The measured interference visibility as a function of photon generation probability (blue data points) and the calculated expected interference visibility for several different radiative decay times (lines).}
\label{Fig2}  
\end{figure}

Once we have injected a hole into the dot with a non-resonant laser pulse, we use two resonant pulses that drive the diagonal transition to create a photon superposed across two time bins. The second resonant pulse requires a higher intensity than the first to compensate for the depletion of the $\ket{h}$ state caused by the first resonant pulse. The outcome is that the photon is equally likely to be measured in each time bin. The capability to produce photons superposed across time bins in this manner has been demonstrated in \cite{lee2018multi}, but here we demonstrate control over the phase difference between the time bins. In principle, this could be done by placing a phase modulator at the output \cite{specht2009phase}, but this introduces losses (at wavelengths of $\sim$ 940 nm, the loss is typically around 3 dB). In our experiments, we show that we can achieve the same result without the associated losses by phase modulating the input resonant driving laser and increasing the laser intensity to account for the loss. 

The pulse sequence used is shown in Figure \ref{Fig2}a. A hole is introduced by a non-resonant pulse and then a non-phase-modulated, two-pulse resonant laser sequence is used to create a photon superposed across two time bins to serve as a reference (pulses $\tau_{1}$ and $\tau_{2}$). Then the hole state is randomised by a second non-resonant pulse (as in \cite{lee2018multi}) to allow the generation of a second photon. The resonant pulses at $\tau_{3}$ and $\tau_{4}$ are used to create a second photon superposed between time bins, but this time the phase modulator is used to modify the phase of the fourth pulse. Directing the light through an unbalanced Michelson interferometer to observe interference between the early and late time bins and time resolving the output allows us to determine the phase difference between the time bins due to the modulator. Figure \ref{Fig2}b shows the result of the experiment for a phase modulation of 0.58$\pi$. The phase change due to the phase modulator is determined by calculating the phase difference between the interference fringes created using the unmodulated reference sequence and modulated sequence (Figure \ref{Fig2}b). The phase modulation has no apparent effect on the interference visibility, with the mean recorded interference visibility being 73.7$\pm$1.1\%.

Taking the phase difference of $\phi = 0$ to represent the $\ket{+}=\frac{1}{\sqrt{2}}(\ket{0}+\ket{1})$ state, using the intensity recorded in the early and late time bins to extract the amplitudes of the $\ket{0}$ and $\ket{1}$ components of the state, and using the interference visibility to help determine the magnitudes of the off-diagonal elements of the density matrix, we plot out the generated states on the Bloch sphere. We determine that we can achieve phase shifts of up to $2.94\pi$ (only shifts below $2\pi$ shown in Figure \ref{Fig2}c). This shows that we can freely control the phase, $\phi$, of the time bin qubit. It is trivial to change the amplitude of the $\ket{0}$ and $\ket{1}$ states by controlling the intensity of the resonant laser pulses, allowing us to conclude that we can use this method to generate any qubit state.

%We note that the phase modulator used was a telecoms wavelength (1.3 $\mu$m) modulator rather than a custom designed modulator for our wavelengths ($\sim$ 940 nm). This resulted in high losses, however, this problem was overcome by simply increasing the laser power before the phase modulator.

This modulation technique could be expanded to higher dimensional states such as those demonstrated in \cite{lee2018multi} in order to create arbitrary time bin encoded qudits. High dimensional single photon qudits have uses in quantum communication protocols \cite{PhysRevA.61.062308, 1367-2630-17-2-022002, sasaki2014practical, takesue2015experimental} and are of interest for quantum computing applications; the use of high dimensional states means that the dimensionality of the Hilbert space needed to describe the states grows faster with photon number than for two dimensional qubit states \cite{schaeff2012scalable}.

\subsection{Coherence and deterministic excitation}
The output photons can be generated in one of two ways: they can be generated coherently by the Raman spin-flip process or by the excitation of the diagonal transition by the resonant laser and the subsequent incoherent decay. In \cite{sweeney2014cavity} and \cite{sun2016measurement}, the authors note that the two processes can be distinguished by their linewidths - the linewidth of the Raman scattered photons is determined by the laser linewidth and the trapped spin coherence time whereas the linewidth of the photons resulting from the incoherent decay is typically broader and has the linewidth of the cavity-enhanced optical transition. The authors of \cite{sweeney2014cavity} observe that, in part due to the cavity enhancement, the Raman process dominates.

%If we assume that the incoherent component is negligible and that the visibility is determined only by the spin coherence time, accounting for the 1.5 ns pulse separation time we extract a spin coherence time of 4.9 ns, which is within the range of previously measured values \cite{}. 

%Although not measured directly, we expect the coherence time of the trapped spin in this dot to fall within the range of 2-10 ns based on previous measurements of dots in this sample and prior work on similar dots. Accounting for the 1.5 ns time bin separation, this give a maximum interference visibility ranging from 47\% to 86\%, meaning that the visibilities shown earlier (averaging 73.7\%) indicate that the Raman process dominates here too.

In analogy with Resonant Rayleigh Scattering, we investigate whether the power of the driving laser increases the proportion of incoherently scattered light \cite{bennett2016cavity}. Using a two level model as in \cite{sun2016measurement}, we expect the ratio between the coherently scattered and incoherently scattered light to be \cite{loudon2000quantum}:
\begin{equation}
\frac{I_{\mathrm{coherent}}}{I_{\mathrm{incoherent}}} = \frac{2\Gamma^{2}}{2\Gamma^{2} + \Omega^{2}},
\label{eqn}
\end{equation}
where $\Gamma=1/T_{1}$ is the radiative decay rate ($T_{1}$ is the radiative decay time) and $\Omega$ is the Rabi frequency.

In our work, we only expect to see interference between the time bins when the photons are produced by Raman scattering. This means that we expect to see a reduction in the interference visibility as the incoherent fraction increases with increasing Rabi frequency.

In order to investigate this effect experimentally, we set the ratio between the first and second resonant laser pulses to be 1:4 in intensity. As the angle of the rotation about the Bloch sphere for a given pulse is proportional to the square root of the power, this means that the second pulse rotates the Bloch vector by twice the angle of the first pulse. We then adjusted the laser power such that the measured intensity of the output light was equal in each time bin and conclude that this means we are driving the $\ket{h}$ to $\ket{T}$ transition with a $\pi /2$ and a $\pi$ pulse for the first and second laser pulses respectively. Therefore, provided the system is in the $\ket{h}$ state initially, this process deterministically creates a photon. Given that this process is limited to a maximum of a single photon per cycle of the pulse sequence (or until a spontaneous spin flip occurs - this is typically on the scale of microseconds \cite{heiss2007observation}, several orders of magnitude longer than the pulse sequence) it does not make sense to consider higher powers than this. We performed the interference measurement at this power and at several lower powers - we have plotted the resulting interference visibilities in Figure \ref{Fig2}d. We observe that the interference visibility decreases at high laser driving powers, but the Raman process still dominates.

%Using resonant pulses of length 400 ps, and assuming decay time of $T_{1}=$250 ps (obtained by taking the standard $\sim$1000 ps decay time seen in non-cavity enhanced dots of this type and dividing it by the Purcell factor), we calculate the Rabi frequency for the brightest pulse in the sequence. Assuming a hole coherence time of 6 ns (which limits maximum visibility to 77.8\%) we plot the expected interference visibility as a function of the probability of creating a photon for per pulse sequence, provided that the initial state is $\ket{h}$) (\ref{Fig2}e).

Using equation \ref{eqn} we can plot out the expected interference visibility as a function of the probability of generating a photon. The maximum achievable visibility is determined by the coherence time of the trapped spin. Estimating a coherence time of $\sim 6$ ns gives reasonable agreement with our results (resulting in a maximum possible visibility of 77.8\% when accounting for the 1.5 ns pulse separation time) and is within the range of previously measured values for the coherence time of a trapped hole spin \cite{sun2016measurement,de2011ultrafast, PhysRevB.89.075316}. 

We then assume that any reduction in the interference visibility below 77.8\% is due to the reduction of the coherent fraction of the scattered light. We use the Rabi frequency of the second (the brightest) pulse to calculate the expected resulting interference visibility as function of the probability of generating a photon. We plot the expected visibilities for several different values of $T_{1}$ and see the $T_{1} = 250$ ps gives good agreement with our experimental results. %This is reasonable as we estimate the Purcell factor to be $\sim 4$ and the typical $T_{1}$ for non-cavity enhanced dots of this type is $\sim 1000$ ps - we would therefore estimate the $T_{1}$ of cavity-enhanced transition to be $\sim 1000/4 = 250$ ps.

We note that for shorter $T_{1}$ times the coherence degrades less with photon generation probability. We anticipate that using higher Purcell factor systems to reduce the radiative decay time would increase the coherent fraction and so increase the interference visibility. It may also be possible to increase the coherent fraction by detuning the cavity and the resonant laser from the $\ket{T}\rightarrow\ket{\bar{h}}$ transition line, as in \cite{sweeney2014cavity}. Moving beyond the simple two-level model in this way may enable the cavity enhancement of the coherent Raman scattered light without directly enhancing the transition.

Our current setup has a relatively small Purcell factor, nevertheless the Raman process dominates at all photon generation probabilities, indicating that this spin-flip Raman scattering technique holds promise for the deterministic generation of arbitrary, d-dimensional, single photon qudits.

\section{Single photon wavelength division multiplexing}

% \begin{figure}[t!]
% \begin{minipage}[h]{0.45\textwidth}
% \includegraphics[width=8 cm]{WDM.png}
% \end{minipage}
% \begin{minipage}[h]{0.45\textwidth}
% \includegraphics[width=8 cm]{g2.png}
% \end{minipage}
% \caption{A time resolved plot and fit of the output light when a red detuned laser is used to drive the Raman transition in time bin 1 and a blue detuned laser is used to drive the transition in time bin 2 when: {\bf a)} The spectral filter is not detuned. {\bf b)} The spectral filter is blue-detuned. {\bf c)} The spectral filter is red detuned.
% {\bf d)} The result of a second-order correlation function measurement on the output light.}
% \label{Fig3}
% \end{figure}

\begin{figure*}[t]
\includegraphics[width=17 cm]{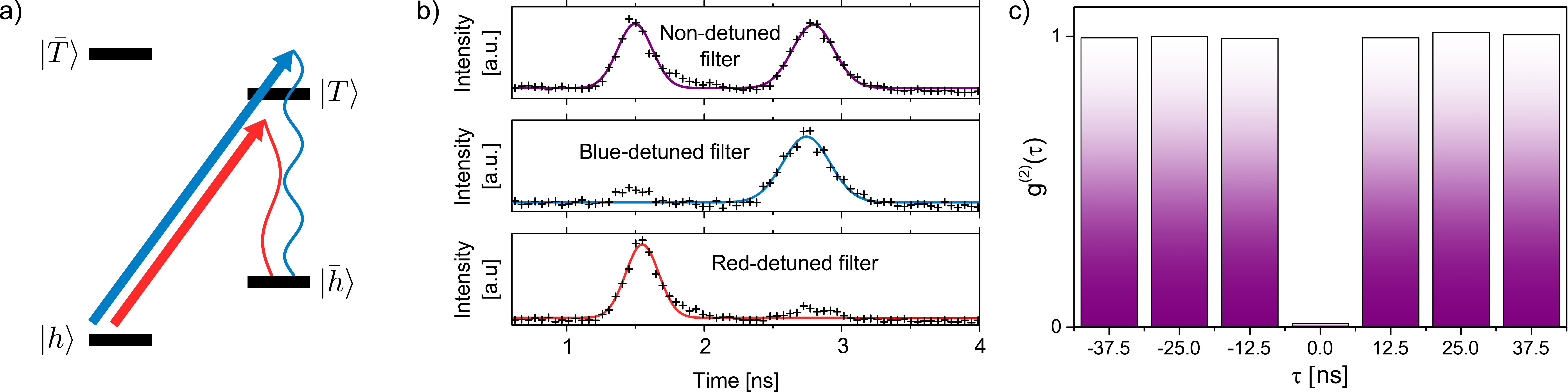}
\caption{{\bf a)} An illustration of driving the Raman transition with a red and a blue detuned laser.
{\bf b)} A time resolved plot and fit of the output light when a red detuned laser is used to drive the Raman transition in time bin 1 and a blue detuned laser is used to drive the transition in time bin 2 when: The spectral filter is not detuned (top); The spectral filter is blue-detuned (middle); The spectral filter is red detuned (bottom).
{\bf c)} The result of a second-order correlation function measurement on the Raman scattered light.}  
\label{Fig3}
\end{figure*}

We now demonstrate the wavelength division multiplexing of a single photon source. Encoding information in different degrees of freedom of single photons is a topic of current interest for quantum communication \cite{wang2015quantum}.

The tuning of the energy of the Raman scattered photons by tuning the driving laser energy has been demonstrated \cite{sweeney2014cavity}. The ability to tune the photon energy is considered important as it enables photons from different sources to be made indistinguishable and so suitable for many quantum communication and computing applications. Here, we use the tunability of Raman scattered photons to encode a different photonic state at two different energies. We use two lasers, red/blue detuned from the diagonal transition by $\sim 10 \ \mu$eV, resulting in a total energy separation of $19.1 \ \mu$eV (illustrated in Figure \ref{Fig3}a). 

In general, a wavelength division multiplexed time bin encoded state generated by this process will have the form:
%\begin{equation}
\begin{align}
\ket{\psi} =& \gamma\ket{\mathrm{red}}(\alpha_{r}\ket{0} + e^{i\phi_{r}}\beta_{r}\ket{1}) \nonumber \\ 
&+ e^{i\phi}\delta\ket{\mathrm{blue}}(\alpha_{b}\ket{0} + e^{i\phi_{b}}\beta_{b}\ket{1}),
\end{align}
%\end{equation}

where $\ket{\mathrm{red}}$ ($\ket{\mathrm{blue}}$) indicates the state generated by the red (blue) detuned laser.

In our experiment, we apply a pulse in time bin 1 with the red detuned laser, encoding the $\ket{0}$ state, and we apply a pulse in time bin 2 with the blue detuned laser, encoding the $\ket{1}$ state. The output state should therefore be:
\begin{equation}
\ket{\psi} = \frac{1}{\sqrt{2}}(\ket{\mathrm{red}}\ket{0} + e^{i\phi}\ket{\mathrm{blue}}\ket{1}).
\end{equation}

Measuring the output shows a roughly equal probability of measuring the $\ket{0}$ or the $\ket{1}$ state (Figure \ref{Fig3}b). However, spectrally filtering the output enables us to recover the $\ket{0}$ state for red detuning and the $\ket{1}$ state for blue detuning (Figure \ref{Fig3}b), demonstrating single photon wavelength division multiplexing. 

As the lasers have no set phase relationship with one another, i.e. we expect the phase, $\phi$, to be random, we do not expect to see interference between the two time bins. In this case, we cannot confirm that this is a coherent superposition state. Future work could create the desired excitation spectrum by modulating and filtering light from a single laser or by phase locking the two driving lasers in order to investigate the possibility of producing coherent superpositions of frequencies for single photons. Finally, we demonstrate that this process does indeed generate single photons by performing a second order correlation function measurement (Figure \ref{Fig3}c). We used time-tagging to remove any influence from photons generated by the non-resonant pulse and considered only the photons generated by driving the Raman transition. This measurement was performed on light generated using two pulses from both detuned lasers. We obtain a value of $g^{(2)}(0)=0.01$, indicating that the output light is primarily composed of single photons, as expected.

\section{Conclusions}
We have demonstrated that we can produce arbitrary single photon time-bin-encoded qubits and that we can in principle do so deterministically, albeit with some loss in coherence. 
%We then suggested some exciting future avenues for research that make use of this demonstration in combination with several other recently demonstrated capabilities. 
We then demonstrated that this cavity-stimulated Raman process can be used to perform wavelength division multiplexing with single photons. In combination with prior work, these results demonstrate the capability to encode large amounts of information with a single photon using the photon energy and high-dimensional, arbitrary time-bin-encoded states.

\section{Methods}

We make use of a QD cavity system that is cooled to 5 K and placed in a 5.5 T Voigt geometry magnetic field. The micropillar cavity is 2.5 $\mu$m in diameter and has a quality factor of $\sim$ 5000. We estimate that the Purcell factor for the long wavelength transition is $\sim 4$ from comparing the intensity of the enhanced an non-enhanced transitions.
Our pulsed non-resonant laser generates light at a wavelength of 850 nm. The amplitude modulation was performed using fibre-coupled \ensuremath{\mathrm{LiNbO_3}} electro-optic amplitude modulators. The phase is controlled with a \ensuremath{\mathrm{LiNbO_3}} electro-optic phase modulator designed for wavelengths of 1.3 $\mu$m; at 940 nm the transmission is $\sim$ 40\%. The output light was measured with silicon APDs.

\section*{Acknowledgments}
The authors acknowledge funding from the EPSRC for MBE system used in the growth of the micropillar cavity.
J. L. gratefully acknowledges financial support from the EPSRC CDT in Photonic Systems Development.
L.W. gratefully acknowledges funding from the EPSRC. 
J.L. and L.W. also gratefully acknowledge financial support Toshiba Research Europe Ltd.
B. V. gratefully acknowledges funding from the European
Union’s Horizon 2020 research and innovation programme
under the Marie Sk\l{}odowska-Curie grant agreement
No. 642688 (SAWtrain).

\section*{Data Access}
The experimental data used to produce the figures in this paper is publicly available at [Information will be made available on Cambridge Data Repository after publication].

\bibliographystyle{apsrev4-1}
\renewcommand{\bibname}{References}
\bibliography{references}

\begin{thebibliography}{28}%
\makeatletter
\providecommand \@ifxundefined [1]{%
 \@ifx{#1\undefined}
}%
\providecommand \@ifnum [1]{%
 \ifnum #1\expandafter \@firstoftwo
 \else \expandafter \@secondoftwo
 \fi
}%
\providecommand \@ifx [1]{%
 \ifx #1\expandafter \@firstoftwo
 \else \expandafter \@secondoftwo
 \fi
}%
\providecommand \natexlab [1]{#1}%
\providecommand \enquote  [1]{``#1''}%
\providecommand \bibnamefont  [1]{#1}%
\providecommand \bibfnamefont [1]{#1}%
\providecommand \citenamefont [1]{#1}%
\providecommand \href@noop [0]{\@secondoftwo}%
\providecommand \href [0]{\begingroup \@sanitize@url \@href}%
\providecommand \@href[1]{\@@startlink{#1}\@@href}%
\providecommand \@@href[1]{\endgroup#1\@@endlink}%
\providecommand \@sanitize@url [0]{\catcode `\\12\catcode `\$12\catcode
  `\&12\catcode `\#12\catcode `\^12\catcode `\_12\catcode `\%12\relax}%
\providecommand \@@startlink[1]{}%
\providecommand \@@endlink[0]{}%
\providecommand \url  [0]{\begingroup\@sanitize@url \@url }%
\providecommand \@url [1]{\endgroup\@href {#1}{\urlprefix }}%
\providecommand \urlprefix  [0]{URL }%
\providecommand \Eprint [0]{\href }%
\providecommand \doibase [0]{http://dx.doi.org/}%
\providecommand \selectlanguage [0]{\@gobble}%
\providecommand \bibinfo  [0]{\@secondoftwo}%
\providecommand \bibfield  [0]{\@secondoftwo}%
\providecommand \translation [1]{[#1]}%
\providecommand \BibitemOpen [0]{}%
\providecommand \bibitemStop [0]{}%
\providecommand \bibitemNoStop [0]{.\EOS\space}%
\providecommand \EOS [0]{\spacefactor3000\relax}%
\providecommand \BibitemShut  [1]{\csname bibitem#1\endcsname}%
\let\auto@bib@innerbib\@empty
%</preamble>
\bibitem [{\citenamefont {Maier}\ \emph {et~al.}(2014)\citenamefont {Maier},
  \citenamefont {Gold}, \citenamefont {Forchel}, \citenamefont {Gregersen},
  \citenamefont {M{\o}rk}, \citenamefont {H{\"o}fling}, \citenamefont
  {Schneider},\ and\ \citenamefont {Kamp}}]{maier2014bright}%
  \BibitemOpen
  \bibfield  {author} {\bibinfo {author} {\bibfnamefont {S.}~\bibnamefont
  {Maier}}, \bibinfo {author} {\bibfnamefont {P.}~\bibnamefont {Gold}},
  \bibinfo {author} {\bibfnamefont {A.}~\bibnamefont {Forchel}}, \bibinfo
  {author} {\bibfnamefont {N.}~\bibnamefont {Gregersen}}, \bibinfo {author}
  {\bibfnamefont {J.}~\bibnamefont {M{\o}rk}}, \bibinfo {author} {\bibfnamefont
  {S.}~\bibnamefont {H{\"o}fling}}, \bibinfo {author} {\bibfnamefont
  {C.}~\bibnamefont {Schneider}}, \ and\ \bibinfo {author} {\bibfnamefont
  {M.}~\bibnamefont {Kamp}},\ }\href@noop {} {\bibfield  {journal} {\bibinfo
  {journal} {Optics express}\ }\textbf {\bibinfo {volume} {22}},\ \bibinfo
  {pages} {8136} (\bibinfo {year} {2014})}\BibitemShut {NoStop}%
\bibitem [{\citenamefont {Pelton}\ \emph {et~al.}(2002)\citenamefont {Pelton},
  \citenamefont {Santori}, \citenamefont {Vuckovi{\'c}}, \citenamefont {Zhang},
  \citenamefont {Solomon}, \citenamefont {Plant},\ and\ \citenamefont
  {Yamamoto}}]{pelton2002efficient}%
  \BibitemOpen
  \bibfield  {author} {\bibinfo {author} {\bibfnamefont {M.}~\bibnamefont
  {Pelton}}, \bibinfo {author} {\bibfnamefont {C.}~\bibnamefont {Santori}},
  \bibinfo {author} {\bibfnamefont {J.}~\bibnamefont {Vuckovi{\'c}}}, \bibinfo
  {author} {\bibfnamefont {B.}~\bibnamefont {Zhang}}, \bibinfo {author}
  {\bibfnamefont {G.~S.}\ \bibnamefont {Solomon}}, \bibinfo {author}
  {\bibfnamefont {J.}~\bibnamefont {Plant}}, \ and\ \bibinfo {author}
  {\bibfnamefont {Y.}~\bibnamefont {Yamamoto}},\ }\href@noop {} {\bibfield
  {journal} {\bibinfo  {journal} {Physical review letters}\ }\textbf {\bibinfo
  {volume} {89}},\ \bibinfo {pages} {233602} (\bibinfo {year}
  {2002})}\BibitemShut {NoStop}%
\bibitem [{\citenamefont {Kuhn}\ \emph {et~al.}(2002)\citenamefont {Kuhn},
  \citenamefont {Hennrich},\ and\ \citenamefont
  {Rempe}}]{kuhn2002deterministic}%
  \BibitemOpen
  \bibfield  {author} {\bibinfo {author} {\bibfnamefont {A.}~\bibnamefont
  {Kuhn}}, \bibinfo {author} {\bibfnamefont {M.}~\bibnamefont {Hennrich}}, \
  and\ \bibinfo {author} {\bibfnamefont {G.}~\bibnamefont {Rempe}},\
  }\href@noop {} {\bibfield  {journal} {\bibinfo  {journal} {Physical review
  letters}\ }\textbf {\bibinfo {volume} {89}},\ \bibinfo {pages} {067901}
  (\bibinfo {year} {2002})}\BibitemShut {NoStop}%
\bibitem [{\citenamefont {Beugnon}\ \emph {et~al.}(2006)\citenamefont
  {Beugnon}, \citenamefont {Jones}, \citenamefont {Dingjan}, \citenamefont
  {Darqui{\'e}}, \citenamefont {Messin}, \citenamefont {Browaeys},\ and\
  \citenamefont {Grangier}}]{beugnon2006quantum}%
  \BibitemOpen
  \bibfield  {author} {\bibinfo {author} {\bibfnamefont {J.}~\bibnamefont
  {Beugnon}}, \bibinfo {author} {\bibfnamefont {M.~P.}\ \bibnamefont {Jones}},
  \bibinfo {author} {\bibfnamefont {J.}~\bibnamefont {Dingjan}}, \bibinfo
  {author} {\bibfnamefont {B.}~\bibnamefont {Darqui{\'e}}}, \bibinfo {author}
  {\bibfnamefont {G.}~\bibnamefont {Messin}}, \bibinfo {author} {\bibfnamefont
  {A.}~\bibnamefont {Browaeys}}, \ and\ \bibinfo {author} {\bibfnamefont
  {P.}~\bibnamefont {Grangier}},\ }\href@noop {} {\bibfield  {journal}
  {\bibinfo  {journal} {Nature}\ }\textbf {\bibinfo {volume} {440}},\ \bibinfo
  {pages} {779} (\bibinfo {year} {2006})}\BibitemShut {NoStop}%
\bibitem [{\citenamefont {Vasilev}\ \emph {et~al.}(2010)\citenamefont
  {Vasilev}, \citenamefont {Ljunggren},\ and\ \citenamefont
  {Kuhn}}]{vasilev2010single}%
  \BibitemOpen
  \bibfield  {author} {\bibinfo {author} {\bibfnamefont {G.~S.}\ \bibnamefont
  {Vasilev}}, \bibinfo {author} {\bibfnamefont {D.}~\bibnamefont {Ljunggren}},
  \ and\ \bibinfo {author} {\bibfnamefont {A.}~\bibnamefont {Kuhn}},\
  }\href@noop {} {\bibfield  {journal} {\bibinfo  {journal} {New Journal of
  Physics}\ }\textbf {\bibinfo {volume} {12}},\ \bibinfo {pages} {063024}
  (\bibinfo {year} {2010})}\BibitemShut {NoStop}%
\bibitem [{\citenamefont {He}\ \emph {et~al.}(2013)\citenamefont {He},
  \citenamefont {He}, \citenamefont {Wei}, \citenamefont {Wu}, \citenamefont
  {Atat{\"u}re}, \citenamefont {Schneider}, \citenamefont {H{\"o}fling},
  \citenamefont {Kamp}, \citenamefont {Lu},\ and\ \citenamefont
  {Pan}}]{he2013demand}%
  \BibitemOpen
  \bibfield  {author} {\bibinfo {author} {\bibfnamefont {Y.-M.}\ \bibnamefont
  {He}}, \bibinfo {author} {\bibfnamefont {Y.}~\bibnamefont {He}}, \bibinfo
  {author} {\bibfnamefont {Y.-J.}\ \bibnamefont {Wei}}, \bibinfo {author}
  {\bibfnamefont {D.}~\bibnamefont {Wu}}, \bibinfo {author} {\bibfnamefont
  {M.}~\bibnamefont {Atat{\"u}re}}, \bibinfo {author} {\bibfnamefont
  {C.}~\bibnamefont {Schneider}}, \bibinfo {author} {\bibfnamefont
  {S.}~\bibnamefont {H{\"o}fling}}, \bibinfo {author} {\bibfnamefont
  {M.}~\bibnamefont {Kamp}}, \bibinfo {author} {\bibfnamefont {C.-Y.}\
  \bibnamefont {Lu}}, \ and\ \bibinfo {author} {\bibfnamefont {J.-W.}\
  \bibnamefont {Pan}},\ }\href@noop {} {\bibfield  {journal} {\bibinfo
  {journal} {Nature nanotechnology}\ }\textbf {\bibinfo {volume} {8}},\
  \bibinfo {pages} {213} (\bibinfo {year} {2013})}\BibitemShut {NoStop}%
\bibitem [{\citenamefont {Matthiesen}\ \emph {et~al.}(2012)\citenamefont
  {Matthiesen}, \citenamefont {Vamivakas},\ and\ \citenamefont
  {Atat{\"u}re}}]{matthiesen2012subnatural}%
  \BibitemOpen
  \bibfield  {author} {\bibinfo {author} {\bibfnamefont {C.}~\bibnamefont
  {Matthiesen}}, \bibinfo {author} {\bibfnamefont {A.~N.}\ \bibnamefont
  {Vamivakas}}, \ and\ \bibinfo {author} {\bibfnamefont {M.}~\bibnamefont
  {Atat{\"u}re}},\ }\href@noop {} {\bibfield  {journal} {\bibinfo  {journal}
  {Physical Review Letters}\ }\textbf {\bibinfo {volume} {108}},\ \bibinfo
  {pages} {093602} (\bibinfo {year} {2012})}\BibitemShut {NoStop}%
\bibitem [{\citenamefont {Matthiesen}\ \emph {et~al.}(2013)\citenamefont
  {Matthiesen}, \citenamefont {Geller}, \citenamefont {Schulte}, \citenamefont
  {Le~Gall}, \citenamefont {Hansom}, \citenamefont {Li}, \citenamefont
  {Hugues}, \citenamefont {Clarke},\ and\ \citenamefont
  {Atat{\"u}re}}]{matthiesen2013phase}%
  \BibitemOpen
  \bibfield  {author} {\bibinfo {author} {\bibfnamefont {C.}~\bibnamefont
  {Matthiesen}}, \bibinfo {author} {\bibfnamefont {M.}~\bibnamefont {Geller}},
  \bibinfo {author} {\bibfnamefont {C.~H.}\ \bibnamefont {Schulte}}, \bibinfo
  {author} {\bibfnamefont {C.}~\bibnamefont {Le~Gall}}, \bibinfo {author}
  {\bibfnamefont {J.}~\bibnamefont {Hansom}}, \bibinfo {author} {\bibfnamefont
  {Z.}~\bibnamefont {Li}}, \bibinfo {author} {\bibfnamefont {M.}~\bibnamefont
  {Hugues}}, \bibinfo {author} {\bibfnamefont {E.}~\bibnamefont {Clarke}}, \
  and\ \bibinfo {author} {\bibfnamefont {M.}~\bibnamefont {Atat{\"u}re}},\
  }\href@noop {} {\bibfield  {journal} {\bibinfo  {journal} {Nature
  communications}\ }\textbf {\bibinfo {volume} {4}},\ \bibinfo {pages} {1600}
  (\bibinfo {year} {2013})}\BibitemShut {NoStop}%
\bibitem [{\citenamefont {Bennett}\ \emph
  {et~al.}(2016{\natexlab{a}})\citenamefont {Bennett}, \citenamefont {Lee},
  \citenamefont {Ellis}, \citenamefont {Meany}, \citenamefont {Murray},
  \citenamefont {Floether}, \citenamefont {Griffths}, \citenamefont {Farrer},
  \citenamefont {Ritchie},\ and\ \citenamefont {Shields}}]{bennett2016cavity}%
  \BibitemOpen
  \bibfield  {author} {\bibinfo {author} {\bibfnamefont {A.~J.}\ \bibnamefont
  {Bennett}}, \bibinfo {author} {\bibfnamefont {J.~P.}\ \bibnamefont {Lee}},
  \bibinfo {author} {\bibfnamefont {D.~J.}\ \bibnamefont {Ellis}}, \bibinfo
  {author} {\bibfnamefont {T.}~\bibnamefont {Meany}}, \bibinfo {author}
  {\bibfnamefont {E.}~\bibnamefont {Murray}}, \bibinfo {author} {\bibfnamefont
  {F.~F.}\ \bibnamefont {Floether}}, \bibinfo {author} {\bibfnamefont {J.~P.}\
  \bibnamefont {Griffths}}, \bibinfo {author} {\bibfnamefont {I.}~\bibnamefont
  {Farrer}}, \bibinfo {author} {\bibfnamefont {D.~A.}\ \bibnamefont {Ritchie}},
  \ and\ \bibinfo {author} {\bibfnamefont {A.~J.}\ \bibnamefont {Shields}},\
  }\href@noop {} {\bibfield  {journal} {\bibinfo  {journal} {Science advances}\
  }\textbf {\bibinfo {volume} {2}},\ \bibinfo {pages} {e1501256} (\bibinfo
  {year} {2016}{\natexlab{a}})}\BibitemShut {NoStop}%
\bibitem [{\citenamefont {Senellart}\ \emph {et~al.}(2017)\citenamefont
  {Senellart}, \citenamefont {Solomon},\ and\ \citenamefont
  {White}}]{senellart2017high}%
  \BibitemOpen
  \bibfield  {author} {\bibinfo {author} {\bibfnamefont {P.}~\bibnamefont
  {Senellart}}, \bibinfo {author} {\bibfnamefont {G.}~\bibnamefont {Solomon}},
  \ and\ \bibinfo {author} {\bibfnamefont {A.}~\bibnamefont {White}},\
  }\href@noop {} {\bibfield  {journal} {\bibinfo  {journal} {Nature
  nanotechnology}\ }\textbf {\bibinfo {volume} {12}},\ \bibinfo {pages} {1026}
  (\bibinfo {year} {2017})}\BibitemShut {NoStop}%
\bibitem [{\citenamefont {Sweeney}\ \emph {et~al.}(2014)\citenamefont
  {Sweeney}, \citenamefont {Carter}, \citenamefont {Bracker}, \citenamefont
  {Kim}, \citenamefont {Kim}, \citenamefont {Yang}, \citenamefont {Vora},
  \citenamefont {Brereton}, \citenamefont {Cleveland},\ and\ \citenamefont
  {Gammon}}]{sweeney2014cavity}%
  \BibitemOpen
  \bibfield  {author} {\bibinfo {author} {\bibfnamefont {T.~M.}\ \bibnamefont
  {Sweeney}}, \bibinfo {author} {\bibfnamefont {S.~G.}\ \bibnamefont {Carter}},
  \bibinfo {author} {\bibfnamefont {A.~S.}\ \bibnamefont {Bracker}}, \bibinfo
  {author} {\bibfnamefont {M.}~\bibnamefont {Kim}}, \bibinfo {author}
  {\bibfnamefont {C.~S.}\ \bibnamefont {Kim}}, \bibinfo {author} {\bibfnamefont
  {L.}~\bibnamefont {Yang}}, \bibinfo {author} {\bibfnamefont {P.~M.}\
  \bibnamefont {Vora}}, \bibinfo {author} {\bibfnamefont {P.~G.}\ \bibnamefont
  {Brereton}}, \bibinfo {author} {\bibfnamefont {E.~R.}\ \bibnamefont
  {Cleveland}}, \ and\ \bibinfo {author} {\bibfnamefont {D.}~\bibnamefont
  {Gammon}},\ }\href@noop {} {\bibfield  {journal} {\bibinfo  {journal} {Nature
  Photonics}\ }\textbf {\bibinfo {volume} {8}},\ \bibinfo {pages} {442}
  (\bibinfo {year} {2014})}\BibitemShut {NoStop}%
\bibitem [{\citenamefont {B{\'e}guin}\ \emph {et~al.}(2017)\citenamefont
  {B{\'e}guin}, \citenamefont {Jahn}, \citenamefont {Wolters}, \citenamefont
  {Reindl}, \citenamefont {Trotta}, \citenamefont {Rastelli}, \citenamefont
  {Ding}, \citenamefont {Huo}, \citenamefont {Schmidt}, \citenamefont
  {Treutlein} \emph {et~al.}}]{beguin2017demand}%
  \BibitemOpen
  \bibfield  {author} {\bibinfo {author} {\bibfnamefont {L.}~\bibnamefont
  {B{\'e}guin}}, \bibinfo {author} {\bibfnamefont {J.-P.}\ \bibnamefont
  {Jahn}}, \bibinfo {author} {\bibfnamefont {J.}~\bibnamefont {Wolters}},
  \bibinfo {author} {\bibfnamefont {M.}~\bibnamefont {Reindl}}, \bibinfo
  {author} {\bibfnamefont {R.}~\bibnamefont {Trotta}}, \bibinfo {author}
  {\bibfnamefont {A.}~\bibnamefont {Rastelli}}, \bibinfo {author}
  {\bibfnamefont {F.}~\bibnamefont {Ding}}, \bibinfo {author} {\bibfnamefont
  {Y.}~\bibnamefont {Huo}}, \bibinfo {author} {\bibfnamefont {O.~G.}\
  \bibnamefont {Schmidt}}, \bibinfo {author} {\bibfnamefont {P.}~\bibnamefont
  {Treutlein}},  \emph {et~al.},\ }\href@noop {} {\bibfield  {journal}
  {\bibinfo  {journal} {arXiv preprint arXiv:1710.02490}\ } (\bibinfo {year}
  {2017})}\BibitemShut {NoStop}%
\bibitem [{\citenamefont {Pursley}\ \emph {et~al.}(2018)\citenamefont
  {Pursley}, \citenamefont {Carter}, \citenamefont {Yakes}, \citenamefont
  {Bracker},\ and\ \citenamefont {Gammon}}]{pursley2018picosecond}%
  \BibitemOpen
  \bibfield  {author} {\bibinfo {author} {\bibfnamefont {B.}~\bibnamefont
  {Pursley}}, \bibinfo {author} {\bibfnamefont {S.}~\bibnamefont {Carter}},
  \bibinfo {author} {\bibfnamefont {M.}~\bibnamefont {Yakes}}, \bibinfo
  {author} {\bibfnamefont {A.}~\bibnamefont {Bracker}}, \ and\ \bibinfo
  {author} {\bibfnamefont {D.}~\bibnamefont {Gammon}},\ }\href@noop {}
  {\bibfield  {journal} {\bibinfo  {journal} {Nature communications}\ }\textbf
  {\bibinfo {volume} {9}},\ \bibinfo {pages} {115} (\bibinfo {year}
  {2018})}\BibitemShut {NoStop}%
\bibitem [{\citenamefont {Lee}\ \emph {et~al.}(2018)\citenamefont {Lee},
  \citenamefont {Bennett}, \citenamefont {Stevenson}, \citenamefont {Ellis},
  \citenamefont {Farrer}, \citenamefont {Ritchie},\ and\ \citenamefont
  {Shields}}]{lee2018multi}%
  \BibitemOpen
  \bibfield  {author} {\bibinfo {author} {\bibfnamefont {J.}~\bibnamefont
  {Lee}}, \bibinfo {author} {\bibfnamefont {A.}~\bibnamefont {Bennett}},
  \bibinfo {author} {\bibfnamefont {R.~M.}\ \bibnamefont {Stevenson}}, \bibinfo
  {author} {\bibfnamefont {D.~J.}\ \bibnamefont {Ellis}}, \bibinfo {author}
  {\bibfnamefont {I.}~\bibnamefont {Farrer}}, \bibinfo {author} {\bibfnamefont
  {D.~A.}\ \bibnamefont {Ritchie}}, \ and\ \bibinfo {author} {\bibfnamefont
  {A.~J.}\ \bibnamefont {Shields}},\ }\href@noop {} {\bibfield  {journal}
  {\bibinfo  {journal} {Quantum Science and Technology}\ } (\bibinfo {year}
  {2018})}\BibitemShut {NoStop}%
\bibitem [{\citenamefont {Warburton}(2013)}]{warburton2013single}%
  \BibitemOpen
  \bibfield  {author} {\bibinfo {author} {\bibfnamefont {R.~J.}\ \bibnamefont
  {Warburton}},\ }\href@noop {} {\bibfield  {journal} {\bibinfo  {journal}
  {Nature materials}\ }\textbf {\bibinfo {volume} {12}},\ \bibinfo {pages}
  {483} (\bibinfo {year} {2013})}\BibitemShut {NoStop}%
\bibitem [{\citenamefont {Bennett}\ \emph
  {et~al.}(2016{\natexlab{b}})\citenamefont {Bennett}, \citenamefont {Lee},
  \citenamefont {Ellis}, \citenamefont {Farrer}, \citenamefont {Ritchie},\ and\
  \citenamefont {Shields}}]{bennett2016semiconductor}%
  \BibitemOpen
  \bibfield  {author} {\bibinfo {author} {\bibfnamefont {A.}~\bibnamefont
  {Bennett}}, \bibinfo {author} {\bibfnamefont {J.}~\bibnamefont {Lee}},
  \bibinfo {author} {\bibfnamefont {D.}~\bibnamefont {Ellis}}, \bibinfo
  {author} {\bibfnamefont {I.}~\bibnamefont {Farrer}}, \bibinfo {author}
  {\bibfnamefont {D.}~\bibnamefont {Ritchie}}, \ and\ \bibinfo {author}
  {\bibfnamefont {A.}~\bibnamefont {Shields}},\ }\href@noop {} {\bibfield
  {journal} {\bibinfo  {journal} {Nature nanotechnology}\ }\textbf {\bibinfo
  {volume} {11}},\ \bibinfo {pages} {857} (\bibinfo {year}
  {2016}{\natexlab{b}})}\BibitemShut {NoStop}%
\bibitem [{\citenamefont {Specht}\ \emph {et~al.}(2009)\citenamefont {Specht},
  \citenamefont {Bochmann}, \citenamefont {M{\"u}cke}, \citenamefont {Weber},
  \citenamefont {Figueroa}, \citenamefont {Moehring},\ and\ \citenamefont
  {Rempe}}]{specht2009phase}%
  \BibitemOpen
  \bibfield  {author} {\bibinfo {author} {\bibfnamefont {H.~P.}\ \bibnamefont
  {Specht}}, \bibinfo {author} {\bibfnamefont {J.}~\bibnamefont {Bochmann}},
  \bibinfo {author} {\bibfnamefont {M.}~\bibnamefont {M{\"u}cke}}, \bibinfo
  {author} {\bibfnamefont {B.}~\bibnamefont {Weber}}, \bibinfo {author}
  {\bibfnamefont {E.}~\bibnamefont {Figueroa}}, \bibinfo {author}
  {\bibfnamefont {D.~L.}\ \bibnamefont {Moehring}}, \ and\ \bibinfo {author}
  {\bibfnamefont {G.}~\bibnamefont {Rempe}},\ }\href@noop {} {\bibfield
  {journal} {\bibinfo  {journal} {Nature Photonics}\ }\textbf {\bibinfo
  {volume} {3}},\ \bibinfo {pages} {469} (\bibinfo {year} {2009})}\BibitemShut
  {NoStop}%
\bibitem [{\citenamefont {Bechmann-Pasquinucci}\ and\ \citenamefont
  {Tittel}(2000)}]{PhysRevA.61.062308}%
  \BibitemOpen
  \bibfield  {author} {\bibinfo {author} {\bibfnamefont {H.}~\bibnamefont
  {Bechmann-Pasquinucci}}\ and\ \bibinfo {author} {\bibfnamefont
  {W.}~\bibnamefont {Tittel}},\ }\href {\doibase 10.1103/PhysRevA.61.062308}
  {\bibfield  {journal} {\bibinfo  {journal} {Phys. Rev. A}\ }\textbf {\bibinfo
  {volume} {61}},\ \bibinfo {pages} {062308} (\bibinfo {year}
  {2000})}\BibitemShut {NoStop}%
\bibitem [{\citenamefont {Zhong}\ \emph {et~al.}(2015)\citenamefont {Zhong},
  \citenamefont {Zhou}, \citenamefont {Horansky}, \citenamefont {Lee},
  \citenamefont {Verma}, \citenamefont {Lita}, \citenamefont {Restelli},
  \citenamefont {Bienfang}, \citenamefont {Mirin}, \citenamefont {Gerrits},
  \citenamefont {Nam}, \citenamefont {Marsili}, \citenamefont {Shaw},
  \citenamefont {Zhang}, \citenamefont {Wang}, \citenamefont {Englund},
  \citenamefont {Wornell}, \citenamefont {Shapiro},\ and\ \citenamefont
  {Wong}}]{1367-2630-17-2-022002}%
  \BibitemOpen
  \bibfield  {author} {\bibinfo {author} {\bibfnamefont {T.}~\bibnamefont
  {Zhong}}, \bibinfo {author} {\bibfnamefont {H.}~\bibnamefont {Zhou}},
  \bibinfo {author} {\bibfnamefont {R.~D.}\ \bibnamefont {Horansky}}, \bibinfo
  {author} {\bibfnamefont {C.}~\bibnamefont {Lee}}, \bibinfo {author}
  {\bibfnamefont {V.~B.}\ \bibnamefont {Verma}}, \bibinfo {author}
  {\bibfnamefont {A.~E.}\ \bibnamefont {Lita}}, \bibinfo {author}
  {\bibfnamefont {A.}~\bibnamefont {Restelli}}, \bibinfo {author}
  {\bibfnamefont {J.~C.}\ \bibnamefont {Bienfang}}, \bibinfo {author}
  {\bibfnamefont {R.~P.}\ \bibnamefont {Mirin}}, \bibinfo {author}
  {\bibfnamefont {T.}~\bibnamefont {Gerrits}}, \bibinfo {author} {\bibfnamefont
  {S.~W.}\ \bibnamefont {Nam}}, \bibinfo {author} {\bibfnamefont
  {F.}~\bibnamefont {Marsili}}, \bibinfo {author} {\bibfnamefont {M.~D.}\
  \bibnamefont {Shaw}}, \bibinfo {author} {\bibfnamefont {Z.}~\bibnamefont
  {Zhang}}, \bibinfo {author} {\bibfnamefont {L.}~\bibnamefont {Wang}},
  \bibinfo {author} {\bibfnamefont {D.}~\bibnamefont {Englund}}, \bibinfo
  {author} {\bibfnamefont {G.~W.}\ \bibnamefont {Wornell}}, \bibinfo {author}
  {\bibfnamefont {J.~H.}\ \bibnamefont {Shapiro}}, \ and\ \bibinfo {author}
  {\bibfnamefont {F.~N.~C.}\ \bibnamefont {Wong}},\ }\href
  {http://stacks.iop.org/1367-2630/17/i=2/a=022002} {\bibfield  {journal}
  {\bibinfo  {journal} {New Journal of Physics}\ }\textbf {\bibinfo {volume}
  {17}},\ \bibinfo {pages} {022002} (\bibinfo {year} {2015})}\BibitemShut
  {NoStop}%
\bibitem [{\citenamefont {Sasaki}\ \emph {et~al.}(2014)\citenamefont {Sasaki},
  \citenamefont {Yamamoto},\ and\ \citenamefont
  {Koashi}}]{sasaki2014practical}%
  \BibitemOpen
  \bibfield  {author} {\bibinfo {author} {\bibfnamefont {T.}~\bibnamefont
  {Sasaki}}, \bibinfo {author} {\bibfnamefont {Y.}~\bibnamefont {Yamamoto}}, \
  and\ \bibinfo {author} {\bibfnamefont {M.}~\bibnamefont {Koashi}},\
  }\href@noop {} {\bibfield  {journal} {\bibinfo  {journal} {Nature}\ }\textbf
  {\bibinfo {volume} {509}},\ \bibinfo {pages} {475} (\bibinfo {year}
  {2014})}\BibitemShut {NoStop}%
\bibitem [{\citenamefont {Takesue}\ \emph {et~al.}(2015)\citenamefont
  {Takesue}, \citenamefont {Sasaki}, \citenamefont {Tamaki},\ and\
  \citenamefont {Koashi}}]{takesue2015experimental}%
  \BibitemOpen
  \bibfield  {author} {\bibinfo {author} {\bibfnamefont {H.}~\bibnamefont
  {Takesue}}, \bibinfo {author} {\bibfnamefont {T.}~\bibnamefont {Sasaki}},
  \bibinfo {author} {\bibfnamefont {K.}~\bibnamefont {Tamaki}}, \ and\ \bibinfo
  {author} {\bibfnamefont {M.}~\bibnamefont {Koashi}},\ }\href@noop {}
  {\bibfield  {journal} {\bibinfo  {journal} {Nature Photonics}\ } (\bibinfo
  {year} {2015})}\BibitemShut {NoStop}%
\bibitem [{\citenamefont {Schaeff}\ \emph {et~al.}(2012)\citenamefont
  {Schaeff}, \citenamefont {Polster}, \citenamefont {Lapkiewicz}, \citenamefont
  {Fickler}, \citenamefont {Ramelow},\ and\ \citenamefont
  {Zeilinger}}]{schaeff2012scalable}%
  \BibitemOpen
  \bibfield  {author} {\bibinfo {author} {\bibfnamefont {C.}~\bibnamefont
  {Schaeff}}, \bibinfo {author} {\bibfnamefont {R.}~\bibnamefont {Polster}},
  \bibinfo {author} {\bibfnamefont {R.}~\bibnamefont {Lapkiewicz}}, \bibinfo
  {author} {\bibfnamefont {R.}~\bibnamefont {Fickler}}, \bibinfo {author}
  {\bibfnamefont {S.}~\bibnamefont {Ramelow}}, \ and\ \bibinfo {author}
  {\bibfnamefont {A.}~\bibnamefont {Zeilinger}},\ }\href@noop {} {\bibfield
  {journal} {\bibinfo  {journal} {Optics Express}\ }\textbf {\bibinfo {volume}
  {20}},\ \bibinfo {pages} {16145} (\bibinfo {year} {2012})}\BibitemShut
  {NoStop}%
\bibitem [{\citenamefont {Sun}\ \emph {et~al.}(2016)\citenamefont {Sun},
  \citenamefont {Delteil}, \citenamefont {Faelt},\ and\ \citenamefont
  {Imamo{\u{g}}lu}}]{sun2016measurement}%
  \BibitemOpen
  \bibfield  {author} {\bibinfo {author} {\bibfnamefont {Z.}~\bibnamefont
  {Sun}}, \bibinfo {author} {\bibfnamefont {A.}~\bibnamefont {Delteil}},
  \bibinfo {author} {\bibfnamefont {S.}~\bibnamefont {Faelt}}, \ and\ \bibinfo
  {author} {\bibfnamefont {A.}~\bibnamefont {Imamo{\u{g}}lu}},\ }\href@noop {}
  {\bibfield  {journal} {\bibinfo  {journal} {Physical Review B}\ }\textbf
  {\bibinfo {volume} {93}},\ \bibinfo {pages} {241302} (\bibinfo {year}
  {2016})}\BibitemShut {NoStop}%
\bibitem [{\citenamefont {Loudon}(2000)}]{loudon2000quantum}%
  \BibitemOpen
  \bibfield  {author} {\bibinfo {author} {\bibfnamefont {R.}~\bibnamefont
  {Loudon}},\ }\href@noop {} {\emph {\bibinfo {title} {The quantum theory of
  light}}}\ (\bibinfo  {publisher} {OUP Oxford},\ \bibinfo {year}
  {2000})\BibitemShut {NoStop}%
\bibitem [{\citenamefont {Heiss}\ \emph {et~al.}(2007)\citenamefont {Heiss},
  \citenamefont {Schaeck}, \citenamefont {Huebl}, \citenamefont {Bichler},
  \citenamefont {Abstreiter}, \citenamefont {Finley}, \citenamefont {Bulaev},\
  and\ \citenamefont {Loss}}]{heiss2007observation}%
  \BibitemOpen
  \bibfield  {author} {\bibinfo {author} {\bibfnamefont {D.}~\bibnamefont
  {Heiss}}, \bibinfo {author} {\bibfnamefont {S.}~\bibnamefont {Schaeck}},
  \bibinfo {author} {\bibfnamefont {H.}~\bibnamefont {Huebl}}, \bibinfo
  {author} {\bibfnamefont {M.}~\bibnamefont {Bichler}}, \bibinfo {author}
  {\bibfnamefont {G.}~\bibnamefont {Abstreiter}}, \bibinfo {author}
  {\bibfnamefont {J.}~\bibnamefont {Finley}}, \bibinfo {author} {\bibfnamefont
  {D.}~\bibnamefont {Bulaev}}, \ and\ \bibinfo {author} {\bibfnamefont
  {D.}~\bibnamefont {Loss}},\ }\href@noop {} {\bibfield  {journal} {\bibinfo
  {journal} {Physical Review B}\ }\textbf {\bibinfo {volume} {76}},\ \bibinfo
  {pages} {241306} (\bibinfo {year} {2007})}\BibitemShut {NoStop}%
\bibitem [{\citenamefont {De~Greve}\ \emph {et~al.}(2011)\citenamefont
  {De~Greve}, \citenamefont {McMahon}, \citenamefont {Press}, \citenamefont
  {Ladd}, \citenamefont {Bisping}, \citenamefont {Schneider}, \citenamefont
  {Kamp}, \citenamefont {Worschech}, \citenamefont {H{\"o}fling}, \citenamefont
  {Forchel} \emph {et~al.}}]{de2011ultrafast}%
  \BibitemOpen
  \bibfield  {author} {\bibinfo {author} {\bibfnamefont {K.}~\bibnamefont
  {De~Greve}}, \bibinfo {author} {\bibfnamefont {P.~L.}\ \bibnamefont
  {McMahon}}, \bibinfo {author} {\bibfnamefont {D.}~\bibnamefont {Press}},
  \bibinfo {author} {\bibfnamefont {T.~D.}\ \bibnamefont {Ladd}}, \bibinfo
  {author} {\bibfnamefont {D.}~\bibnamefont {Bisping}}, \bibinfo {author}
  {\bibfnamefont {C.}~\bibnamefont {Schneider}}, \bibinfo {author}
  {\bibfnamefont {M.}~\bibnamefont {Kamp}}, \bibinfo {author} {\bibfnamefont
  {L.}~\bibnamefont {Worschech}}, \bibinfo {author} {\bibfnamefont
  {S.}~\bibnamefont {H{\"o}fling}}, \bibinfo {author} {\bibfnamefont
  {A.}~\bibnamefont {Forchel}},  \emph {et~al.},\ }\href@noop {} {\bibfield
  {journal} {\bibinfo  {journal} {Nature Physics}\ }\textbf {\bibinfo {volume}
  {7}},\ \bibinfo {pages} {872} (\bibinfo {year} {2011})}\BibitemShut {NoStop}%
\bibitem [{\citenamefont {Carter}\ \emph {et~al.}(2014)\citenamefont {Carter},
  \citenamefont {Economou}, \citenamefont {Greilich}, \citenamefont {Barnes},
  \citenamefont {Sweeney}, \citenamefont {Bracker},\ and\ \citenamefont
  {Gammon}}]{PhysRevB.89.075316}%
  \BibitemOpen
  \bibfield  {author} {\bibinfo {author} {\bibfnamefont {S.~G.}\ \bibnamefont
  {Carter}}, \bibinfo {author} {\bibfnamefont {S.~E.}\ \bibnamefont
  {Economou}}, \bibinfo {author} {\bibfnamefont {A.}~\bibnamefont {Greilich}},
  \bibinfo {author} {\bibfnamefont {E.}~\bibnamefont {Barnes}}, \bibinfo
  {author} {\bibfnamefont {T.}~\bibnamefont {Sweeney}}, \bibinfo {author}
  {\bibfnamefont {A.~S.}\ \bibnamefont {Bracker}}, \ and\ \bibinfo {author}
  {\bibfnamefont {D.}~\bibnamefont {Gammon}},\ }\href {\doibase
  10.1103/PhysRevB.89.075316} {\bibfield  {journal} {\bibinfo  {journal} {Phys.
  Rev. B}\ }\textbf {\bibinfo {volume} {89}},\ \bibinfo {pages} {075316}
  (\bibinfo {year} {2014})}\BibitemShut {NoStop}%
\bibitem [{\citenamefont {Wang}\ \emph {et~al.}(2015)\citenamefont {Wang},
  \citenamefont {Cai}, \citenamefont {Su}, \citenamefont {Chen}, \citenamefont
  {Wu}, \citenamefont {Li}, \citenamefont {Liu}, \citenamefont {Lu},\ and\
  \citenamefont {Pan}}]{wang2015quantum}%
  \BibitemOpen
  \bibfield  {author} {\bibinfo {author} {\bibfnamefont {X.-L.}\ \bibnamefont
  {Wang}}, \bibinfo {author} {\bibfnamefont {X.-D.}\ \bibnamefont {Cai}},
  \bibinfo {author} {\bibfnamefont {Z.-E.}\ \bibnamefont {Su}}, \bibinfo
  {author} {\bibfnamefont {M.-C.}\ \bibnamefont {Chen}}, \bibinfo {author}
  {\bibfnamefont {D.}~\bibnamefont {Wu}}, \bibinfo {author} {\bibfnamefont
  {L.}~\bibnamefont {Li}}, \bibinfo {author} {\bibfnamefont {N.-L.}\
  \bibnamefont {Liu}}, \bibinfo {author} {\bibfnamefont {C.-Y.}\ \bibnamefont
  {Lu}}, \ and\ \bibinfo {author} {\bibfnamefont {J.-W.}\ \bibnamefont {Pan}},\
  }\href@noop {} {\bibfield  {journal} {\bibinfo  {journal} {Nature}\ }\textbf
  {\bibinfo {volume} {518}},\ \bibinfo {pages} {516} (\bibinfo {year}
  {2015})}\BibitemShut {NoStop}%
\end{thebibliography}%

\end{document}